# Simulations Predict Improved Valve Performance Without Direct Leaflet Intervention After Neonatal Truncus Arteriosus Repair

Karoline-Marie Bornemann, PhD[1,2], Perry S. Choi, MD[2,3,4], Jay Huber, MAB[3],

Alexander K. Reed, MD[3,4], Amit Sharir[3,4], Shiraz A. Maskatia, MD[1,2,4],

Alison L. Marsden, PhD[1,2,4,5,6], Michael R. Ma, MD[2,3,4], Alexander D. Kaiser, PhD[2,3,4]

[1]*Stanford University, Department of Pediatrics, Division of Pediatric Cardiology, Palo Alto, CA*

[2]*Stanford University, Maternal & Child Health Research Institute, Palo Alto, CA*

[3]*Stanford University, Department of Cardiothoracic Surgery, Palo Alto, CA*

[4]*Stanford University, Cardiovascular Institute, Palo Alto, CA*

[5]*Stanford University, Department of Bioengineering, Palo Alto, CA*

[6]*Stanford University, Institute for Computational & Mathematical Engineering, Palo Alto, CA*

**Disclosure Statement**

The authors report no conflicts of interest.

**Funding Statement**

KMB and ADK were supported in part by NIH grant K25HL175208, AHA Career Development Award 24CDA1272816 and Stanford Maternal and Child Health Research Institute. AKR was funded by NIH grant R38 HL143615.

**Institutional Review Board (IRB) Approval**

IRB number: 39377, IRB approval date: June 17, 2025

**ABSTRACT**

**Objectives:** Truncus arteriosus is a rare and severe congenital heart pathology. Quadricuspid valve morphology occurs in a quarter of all truncal patients and is linked to valve regurgitation and increased risk of re-operation. It remains unclear how sudden hemodynamic changes during truncal repair alter valve performance. This study simulated pre- and postoperative conditions in a neonatal truncal patient to investigate quadricuspid valve performance before and after truncal repair without leaflet intervention. We hypothesize that valve performance before and after truncal repair can be predicted with in-silico simulations, matching clinical in-vivo imaging and identifying fundamental mechanisms how hemodynamic changes after truncal repair will reduce valve regurgitation without direct intervention.

**Methods:** Pre- and postoperative computed tomography images of a neonatal patient with truncus arteriosus and quadricuspid valve morphology were segmented to extract the patient-specific geometry. Free edge length and geometric height from the patient's echocardiogram were used to model the quadricuspid valve. For the preoperative condition, left and right ventricular pressures were set equal modeling an unrestricted ventricular septal defect. Systemic and pulmonary resistances were tuned based on the patient's Qp:Qs ratio. For the postoperative condition, boundary conditions were modified to mimic patient-specific hemodynamic conditions after truncal repair. Quadricuspid valve performance was evaluated based on valve regurgitation, leaflet motion, and hemodynamics.

**Results:** The preoperative simulation confirmed mild valve regurgitation seen in the preoperative echocardiogram. Despite leaflet symmetry, interaction between asymmetric flow and the surrounding vessel resulted in asymmetric opening and closing. Poor central leaflet coaptation led to a central regurgitant jet toward the septum. Despite no change in leaflet or truncal root morphology, altered postoperative hemodynamic conditions improved leaflet coaptation and eliminated regurgitation, as seen in postoperative imaging. Skewed inflow from the left ventricle led to asymmetric leaflet motion and aortic flow inclined against the greater curvature.

**Conclusions:** The presented modeling approach based on segmented patient data and tuned patient-specific hemodynamics reproduced pre- and postoperative valve performance in line with in-vivo imaging and identified mechanisms improving leaflet coaptation after truncal repair. In this first computational study of a truncal patient, truncal repair led to elimination of valve regurgitation due to enhanced diastolic central leaflet coaptation facilitated by favorable interaction between blood flow and the surrounding vessel. Thus, altered postoperative hemodynamic conditions after truncal repair may improve valve performance without direct leaflet intervention.

**INTRODUCTION**

Truncus arteriosus (TA) is a rare and severe congenital heart disease (CHD) accounting for 1% to 4% of all CHD.[1,2] Early mortality rates of TA patients range from 3 to 20% with 20-year survival rate of 76.8% after truncus arteriosus repair.[3]

During embryogenesis, the separation of the embryologic arterial truncus and semilunar valves fails leading to one common truncus with one truncal valve.[4] In TA patients, blood from both ventricles enters the truncal valve overriding a large ventricular septal defect (VSD).

Patients with TA require surgical repair optimally within the neonatal stage as delayed repair is associated with detrimental short- and long-term outcomes.[7,8] While highly individualized and dependent on the patient-specific TA anatomy, the division of systemic and pulmonary circulation is generally performed by separating the pulmonary arteries from the common truncus and reconnecting them to the right ventricle (RV) via a valved RV-PA conduit. Often, truncal valve pathology necessitates intervention.[10] However, risk for reintervention remains high.[10]

The usually dysplastic truncal valve features thickened leaflets and may be unicuspid, bicuspid, trileaflet or quadricuspid. While trileaflet truncal valves are the most common truncal valve phenotype with 66% incidence, quadricuspid valve morphology is observed in 25% of all TA patients, of which a third of these require immediate surgical valve intervention.[10] Due to high rates of moderate to severe valve regurgitation, patients with quadricuspid valve morphology are also at higher risk for re-intervention during early childhood.[10]

To avoid future valve complications, surgeons frequently prefer to avoid any direct leaflet intervention. However, it is unclear how changes in hemodynamics following truncal repair affect the truncal valve performance itself. By separating the pulmonary and systemic circulation during truncal repair, the total flow through the valve is substantially decreased and the truncal valve is subjected to the higher pressure levels of the systemic circulation.

In this work, pre- and postoperative conditions in a neonatal truncal patient with regurgitant quadricuspid valve morphology to investigate truncal valve performance before and after truncal repair without direct leaflet intervention were simulated. Performing fluid-structure interaction simulation on the patient-specific pre- and postoperative geometry, we evaluated truncal valve performance based on valve regurgitation, leaflet motion, and hemodynamics in pre- and postoperative conditions.

## MATERIALS AND METHODS

### Simulation Methods

A neonatal patient with truncus arteriosus (Collett and Edwards type II) and quadricuspid valve morphology was selected based on the availability of pre- and postoperative computed tomography (CT) and echocardiography imaging. The effect of hemodynamic changes on valve performance could be isolated as the patient's truncal valve was not surgically modified during truncus arteriosus repair. Preoperative (patient age: 2 days) and postoperative (patient age: 1 month) CT images were segmented using SimVascular.[11] The computational domain of the patient-specific preoperative geometry includes portions of the left (LV) and right ventricle (RV) connected by a large ventricular septal defect (VSD) and a right-sided truncus with left (LPA) and right pulmonary arteries (RPA) originating approximately at the sinotubular junction (STJ). The postoperative geometry included LV and right-sided aortic arch.

The quadricuspid valve was modeled with four symmetric leaflets matching the patient's free edge length (FEL) and geometric height (GH) measured from their echocardiogram before and after truncal surgery. The preoperative echocardiogram revealed an annular diameter of $d_A$ = 7.25 mm, a free edge length of FEL = 7.56 mm and a geometric height of GH = 6.5 mm. Postoperative echocardiography imaging showed an annular diameter of $d_A$ = 8.1 mm, free edge length of FEL = 8.95 mm, and geometric height of GH = 6.5 mm. The leaflets were attached to a rigid symmetrical scaffold with four commissures which were oriented according to CT images. Valve orientation was validated by the direction of the regurgitant jet in preoperative echocardiography imaging (Figure 5). The valve thickness in both pre- and postoperative configuration was set to 0.5 mm according to clinically reported "thickened" leaflets in this patient. The truncal valve leaflet shape were designed according to an elasticity-based design approach.[13–18] Assuming that the leaflet tension supports a prescribed pressure load, we derived partial differential equations describing its mechanical equilibrium. Their solution predicted a loaded leaflet configuration of the closed valve. From this pressurized configuration, a constitutive law for the valve with scaled membrane stiffnesses was derived to achieve the predicted tensions. This computational framework for valve generation has shown excellent agreement with ex-vivo testing[19] and in-vitro experiments[17] in previous studies.

Fluid-structure interaction simulations were performed with the Immersed Boundary Method which is commonly applied for numerical modeling problems involving heart valves.[12,19,20] In this study, the open-source software Immersed Boundary Adaptive Mesh Refinement (IBAMR) was used.[21]

Boundary conditions were tuned based on patient-specific clinical data CT and echocardiography imaging as well as patient hemodynamics before and after truncus arteriosus repair. In the preoperative configuration, the heart rate was 146 bpm resulting in a cardiac cycle duration of 0.411 s. Left and right ventricular pressures were assumed equal due to a large, unrestricted VSD. The left ventricular pressure was prescribed based on a time-varying elastance model for a stroke volume of 2.4 ml with neonatal end-systolic ventricular elastance.[22–24] A Resistance-Capacitance-Resistance (RCR) boundary condition[25] at the distal truncal arch was tuned to match patient-specific systolic/diastolic pressures of 60/24 mmHg. The patient-specific preoperative Qp:Qs ratio was estimated to be between 3:1 and 4:1. This ratio was modeled by adjusting the ratio of physiological values for neonatal systemic vascular resistance (SVR)[25] and pulmonary vascular resistance (PVR)[26], resulting in a Qp:Qs of 3.2:1 in our simulations. In the postoperative configuration, the heart rate was 157 bpm, equal to a cardiac cycle duration of 0.382 s. The left ventricular stroke volume was 2.0 ml. The distal aortic arch showed systolic/diastolic pressures of 76/38 mmHg.

The Institutional Review Board of the Stanford University approved the study protocol and publication of data. Patient written consent for the publication of the study data was waived by the Institutional Review Board for use of anonymized, retrospectively acquired data (#39377, June 17, 2025).

**Evaluation Metrics**

Quadricuspid valve performance was evaluated based on regurgitant fraction, leaflet motion, and hemodynamics. Valve regurgitation is evaluated based on regurgitant fraction.[27] Leaflet kinematics are assessed dynamically during one cardiac cycle. During diastole, we assess efficiency of leaflet coaptation by evaluating leaflet coaptation height[28], regurgitant jet width relative to annular diameter measured at the leaflet tip and maximum regurgitant jet velocity.

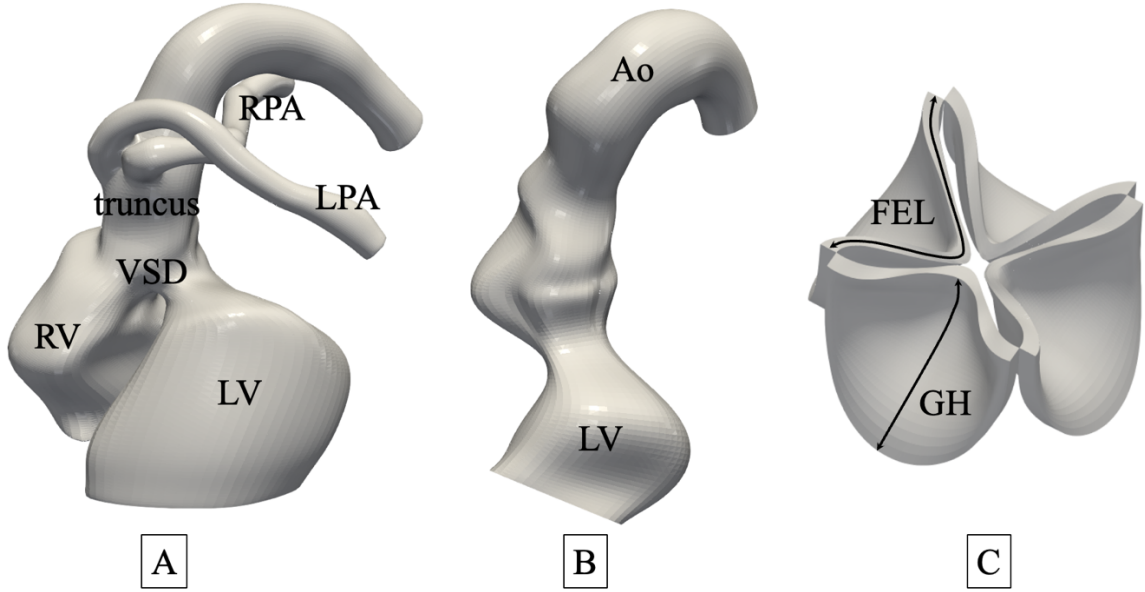

Figure 1: (A) Preoperative geometry: VSD between left and right ventricle, pulmonary arteries originating above the truncal valve, right-sided aortic arch. (B) Postoperative geometry: closed VSD, detached pulmonary arteries. (C) Quadricuspid valve morphology with leaflet geometric height (GH) and leaflet free edge length (FEL).

Table 1: Pre- and postoperative clinical parameters

|  | PREOPERATIVE | POSTOPERATIVE |
|---|---|---|
| Heart rate [bpm] | 146 | 157 |
| Cycle duration [s] | 0.411 | 0.382 |
| Left ventricular stroke volume [ml] | 2.4 | 2.0 |
| Systolic / diastolic pressure [mmHg] | 60 / 24 | 76 / 38 |
| Valve annular diameter ($d_A$) [mm] | 7.25 | 8.1 |
| Leaflet geometric height (GH) [mm] | 6.5 | 6.5 |
| Leaflet free edge length (FEL) [mm] | 7.56 | 8.95 |

**RESULTS**

The preoperative simulation confirmed mild valve regurgitation found in the preoperative echocardiogram. Despite leaflet symmetry, interaction between asymmetric flow and the surrounding vessel resulted in asymmetric opening and closing. Poor central leaflet coaptation led to a central regurgitant jet across the VSD orifice impacting the septum. Despite no change in leaflet or truncal root morphology, altered postoperative hemodynamic conditions revealed better leaflet coaptation, eliminating regurgitation. Skewed inflow from the LV led to asymmetric leaflet kinematics and aortic flow inclined against the greater curvature.

Figure 2 shows the velocity magnitude in slices through left and right ventricle as well as truncus and pulmonary arteries for the preoperative configuration. Starting from the end of the isovolumetric contraction to the end of the diastolic filling phase, velocity magnitude and leaflet kinematics is displayed over time. Despite the symmetric geometry of the quadricuspid valve, upstream and downstream pressure fields led to an asymmetric diastolic configuration. During systolic acceleration, the valve opened asymmetrically and showed random oscillatory leaflet motion of small amplitude during systole. The open valve configuration showed excessive bulging of the thickened leaflets and a resulting S-shape of the leaflet free edge. The systolic jet was inclined towards the greater curvature and created forward flow in truncus and pulmonary arteries. During valve closure, the posterior leaflet located at the inner curvature of the truncus closed first, while its neighboring leaflet, located above the left ventricle, closed last. A strong interaction between valve closure and the surrounding asymmetric flow field was observed. During diastole, poor central leaflet coaptation in the valve center led to a narrow regurgitant jet (width approximately 10% of the annular diameter) directed across the VSD and impacting the upper part of the septum. The leaflet coaptation height near the commissures was 1.3 mm, equaling 20% of the geometric leaflet height. The maximum velocity of the regurgitant jet is 282.5 cm/s. As evident in Figure 4, pressure and wave forms of the preoperative configuration revealed that regurgitation is fed from the truncus while forward flow persists in the pulmonary arteries during the entire cardiac cycle.

As summarized in Table 2, regurgitation in the preoperative configuration was classified as mild (regurgitant fraction < 30%, jet width to annular diameter < 25%) which is validated by echocardiography data and clinical patient reports. Moreover, the direction and intensity of the regurgitant jet can be validated by echocardiography as seen in Figure 5.

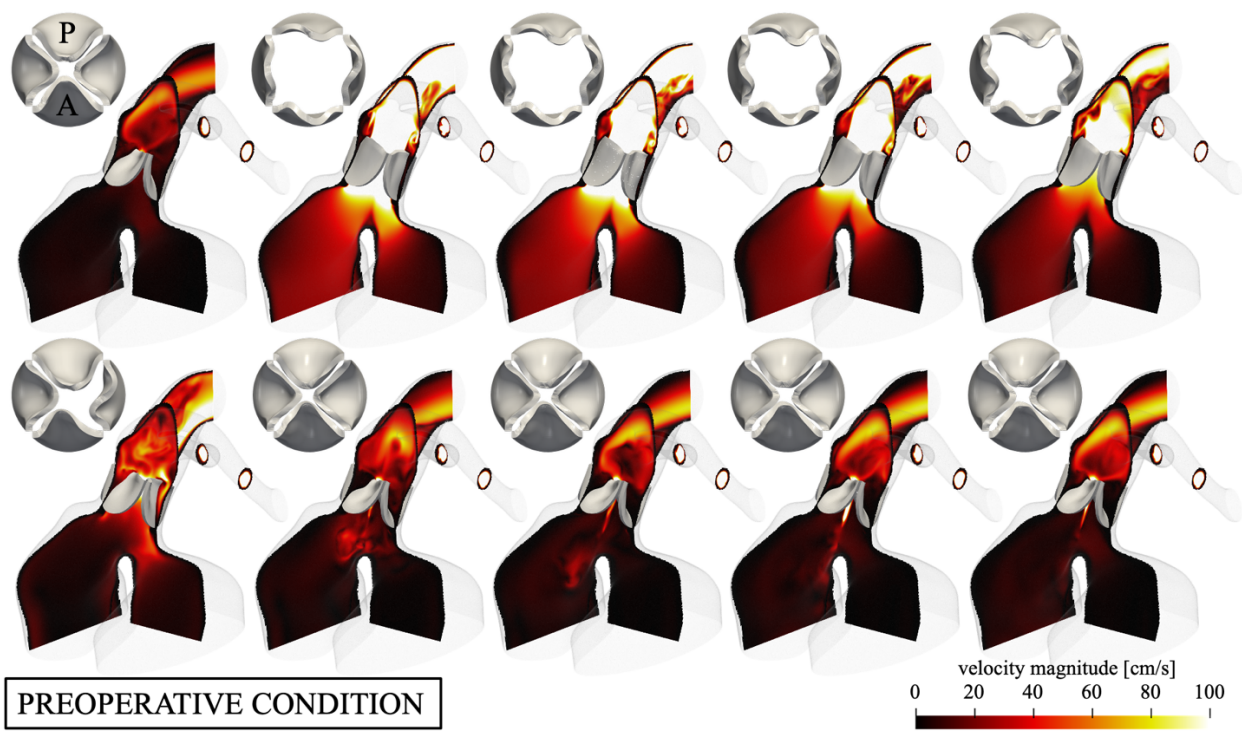

Figure 2: Preoperative flow fields from end of isovolumetric contraction to end of diastolic filling phase. Velocity magnitude at different slices and opening and closing kinematics of the quadricuspid truncal valve.

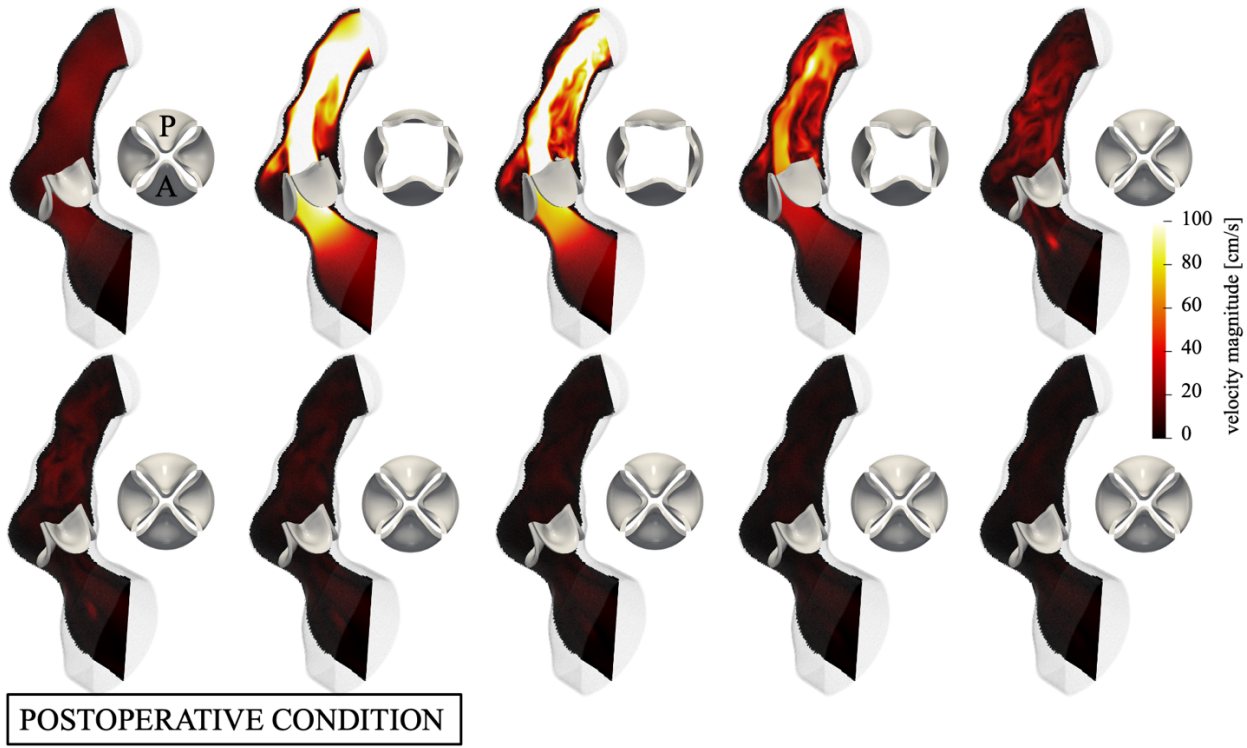

Figure 3: Postoperative flow fields from end of isovolumetric contraction to end of diastolic filling phase. Velocity magnitude at different slices and opening and closing kinematics of the quadricuspid truncal valve. (Reoriented compared to Figure 2 to display better view.)

Figure 3 shows the flow fields and leaflet kinematics after truncal repair. The closed leaflet configuration before opening was much more symmetric than the preoperative configuration. During systole, the valve leaflets remained in an asymmetric open configuration with some leaflets exhibiting random, oscillatory motion of small amplitude. The aortic jet impacted the wall immediately downstream of the valve leading to backflow towards the sinus portions and inclination towards the greater curvature. During systolic deceleration, the jet broke down into smaller vortices, and the adverse pressure gradient led to valve closure. No regurgitation existed during diastole, with a small regurgitant fraction resulting from an initial closing transient (Table 2). The leaflets properly coapted in the valve center in contrast to the preoperative configuration and the leaflet coaptation height increased from 1.3 mm to 2 mm (30.8% of leaflet geometric height).

Table 2: Regurgitation parameters during diastole indicating mild regurgitation (regurgitant fraction < 30%, jet width to annular diameter < 25%) in preoperative condition.

|  | PREOPERATIVE | POSTOPERATIVE |
|---|---|---|
| Regurgitant fraction [%] | 12.3 | 6.5 |
| Leaflet coaptation height [mm] | 1.3 | 2.0 |
| Maximum velocity regurgitant jet [cm/s] | 282.5 | 0.0 |
| Jet width to annular diameter [%] | 10 | 0.0 |

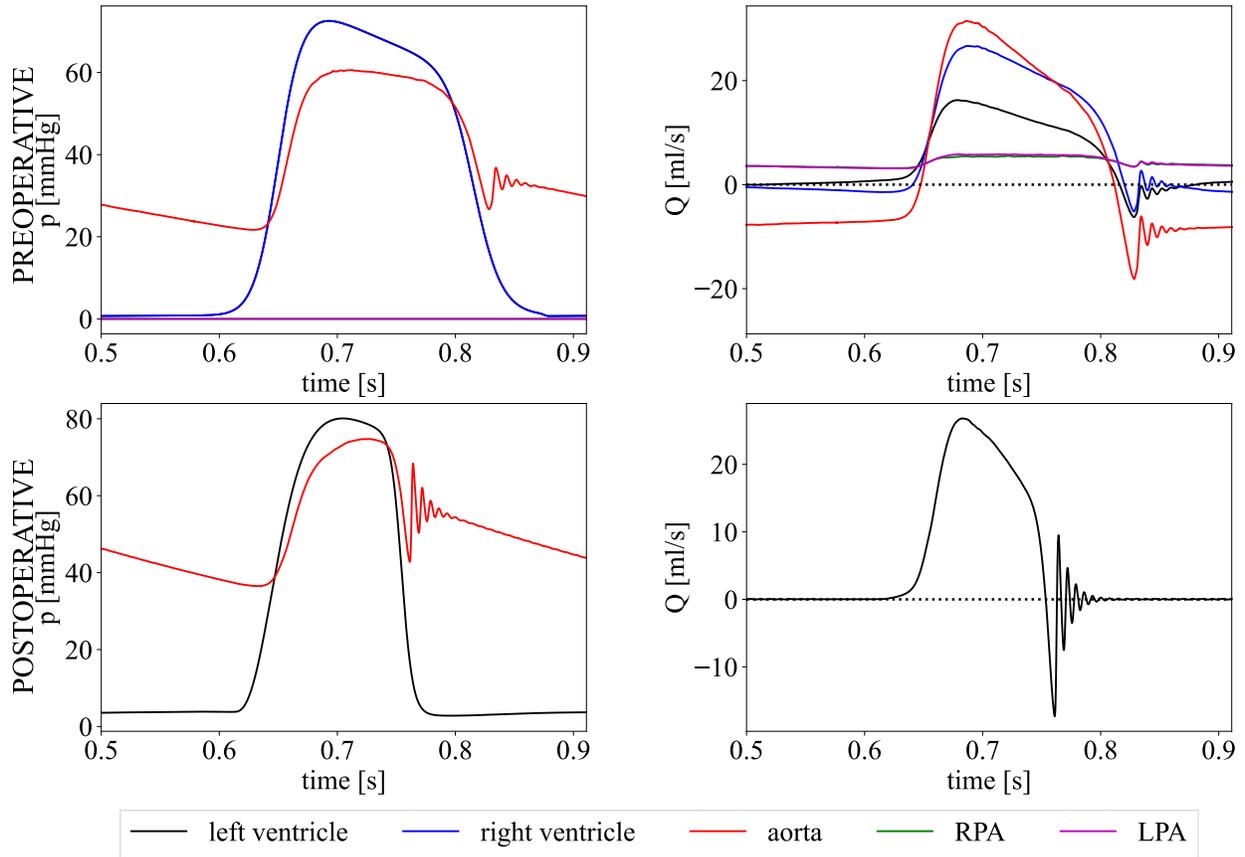

Figure 4: Preoperative and postoperative pressure and flow waveforms.

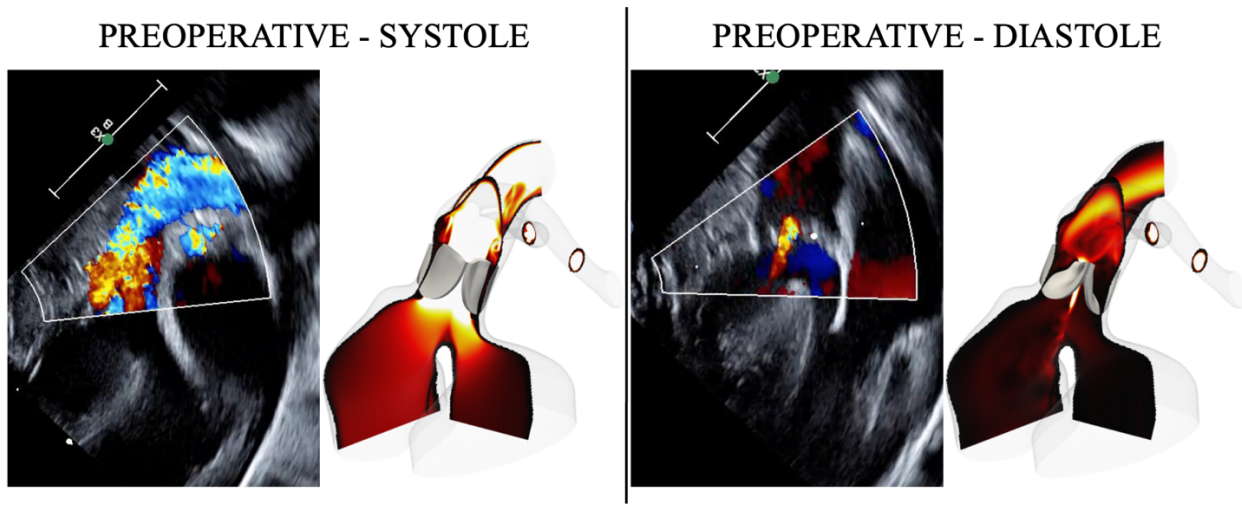

Figure 5: Validation of numerical results with echocardiography images confirmed that poor leaflet coaptation during diastole led to central regurgitant jet across VSD orifice impacting the septum towards the right ventricle. Echocardiography confirmed non-zero flow velocities in truncus and pulmonary arteries during diastole in consistency with numerical results.

**DISCUSSION**

This first patient-specific fluid-structure interaction simulation of a neonatal patient with truncus arteriosus and a quadricuspid valve indicated that changes in hemodynamic conditions caused by truncal repair might eliminate valve regurgitation without direct leaflet intervention. Numerical results reproduced central clinical findings by showing mild valve regurgitation found in the preoperative echocardiogram and clinical patient reports. No regurgitation was found in the postoperative echocardiogram and simulations reproduced this trend.

In line with clinical imaging of quadricuspid valves, numerical simulations confirmed the main mechanism of valve regurgitation of a symmetric quadricuspid valve as poor central leaflet coaptation in the valve center. One reason for poor central coaptation could be purely geometrical. If the leaflet free edge length is considerably shorter than the annular diameter, the center of the leaflet free edge cannot reach the valve center, hence leading to insufficient coaptation. However, the free edge length to annular diameter ratio was similar in the simulated pre- and postoperative configuration, suggesting that another mechanism played a role in the occurrence of valve regurgitation. Given the significant change in hemodynamic conditions and surrounding vessel geometry before and after truncal repair, the interaction between surrounding flow field, valve and surrounding vessels seemed to play a crucial role in valve performance. In the preoperative configuration, the truncus showed a tubular shape with little appearance of a defined sinus of Valsalva, restricting outward leaflet motion and leading to excessive leaflet buckling during systole. The surrounding flow field was highly asymmetric as blood flow entered from both left and right ventricle through a VSD and exited through three outlets (truncus, LPA, RPA). The pulmonary arteries were located in close proximity, contributing to the asymmetry of the downstream flow field. As a result, we observed highly asymmetric open and closed leaflet configurations with poor central coaptation.

In the postoperative configuration, however, leaflet kinematics were much more symmetrical, which might be related to two factors: (a) less complex surrounding vessel geometry and (b) definition and size of the sinus portions. First, the VSD was closed, and the pulmonary arteries were disconnected from the truncus during truncal surgery. Hence, blood flow entered the valve from only one ventricle and exists the valve through one great artery. Moreover, the previously tubular truncus was slightly reshaped through the disconnection of the pulmonary arteries. In contrast to the preoperative CT, more pronounced sinus portions were visible in the postoperative CT, extending the space surrounding the valve. This could potentially contribute to a more symmetric opening of the valve as the leaflets were less restricted in their opening motion. Reshaping the truncus towards an aortic root geometry with sinus portions and sinotubular junction could therefore potentially benefit valve performance.

As the valve model was mounted in a rigid ring and inserted into a rigid vessel geometry, flexibility of the surrounding tissue might slightly alter hemodynamics but is unlikely to change general mechanisms.

The central mechanisms resulting in improved postoperative valve performance may also be applicable to aortic and pulmonary valves in a broader range of congenital pathologies. In further studies, we will extend the patient cohort to generalize findings across multiple types of truncus arteriosus and truncal valves with the goal of predicting postoperative valve performance even before truncal surgery to guide surgical planning.

**CONCLUSIONS**

The presented modeling approach based on segmented patient data and tuned patient-specific hemodynamics reproduced pre- and postoperative valve performance in line with in-vivo imaging and identified mechanisms improving leaflet coaptation after truncal repair. In this first computational study of a truncal patient, truncal repair led to elimination of valve regurgitation due to enhanced diastolic central leaflet coaptation facilitated by favorable interaction between blood flow and the surrounding vessel. Thus, altered postoperative hemodynamic conditions after truncal repair may improve valve performance without direct leaflet intervention.